\documentclass[aps,pra,reprint,superscriptaddress]{revtex4-2}
\usepackage{float}
\usepackage{amsmath}
\usepackage{amsfonts}
\usepackage{dcolumn}
\usepackage{bm}
\usepackage{lipsum,mwe}
\usepackage{graphicx}
\usepackage{epsfig}
\usepackage{epstopdf}
\usepackage{graphicx}
\usepackage{amsfonts}
\usepackage{amssymb}
\usepackage{mathrsfs}
\usepackage{color}
\usepackage{appendix}
\usepackage{multirow}
\usepackage{threeparttable}
\usepackage{verbatim}
\usepackage{graphicx,amsmath}
\usepackage{color,xcolor}
\usepackage[bookmarks=false]{hyperref}
\hypersetup{colorlinks=true,citecolor=blue,linkcolor=blue,urlcolor=blue,pdfstartview=FitH,bookmarksopen=true}

\begin{document}
\title{Enhancement of quantum sensing in a dissipatively coupled two-mode system}
\author{Hao-Wen Zhang}
\affiliation{School of Physics, Zhengzhou University, Zhengzhou 450001, China}
\author{Dong-Yang Wang}\email[]{dywang@zzu.edu.cn}
\affiliation{School of Physics, Zhengzhou University, Zhengzhou 450001, China}
\affiliation{Institute of Quantum Science and Technology, Yanbian University, Yanji, Jilin 133002, China}
\author{Cheng-Hua Bai}
\affiliation{School of Semiconductor and Physics, North University of China, Taiyuan, Shanxi 030051, China}
\author{Tian-Xiang Lu}
\affiliation{College of Physics and Electronic Information, Gannan Normal University, Ganzhou 341000, China}
\author{Shi-Lei Su}\email[]{slsu@zzu.edu.cn}
\affiliation{School of Physics, Zhengzhou University, Zhengzhou 450001, China}
\date{\today}

\begin{abstract}
Quantum sensing near exceptional points (EPs) in non-Hermitian systems has shown promising sensitivity enhancements. However, practical applications are often hindered by structural complexity and strict parameter constraints. In this work, we introduce a simplified anti-parity-time (anti-PT) symmetric platform consisting of two independently cavities, which are indirectly coupled to each other by a shared dissipative environment. We demonstrate a significantly enhanced sensing response at the EPs compared to non-EP configurations. This improvement is attributed to the dominant second-order term in the Laurent series expansion of the eigenvalue response to external perturbations- a characteristic feature of higher-order singularities at EPs. This mechanism not only reinforces the foundation for sensitivity enhancement but also offers a structurally compact and robust strategy for quantum sensing. Our results underscore the potential of anti-PT symmetric systems in enabling high-precision sensing technologies and bridging non-Hermitian physics with scalable photonic device platforms.
\end{abstract}

\maketitle

\section{INTRODUCTION}\label{sec1}
Quantum sensing~\cite{DeMille2024,Pirandola2018,MONTENEGRO20251} has emerged as a transformative approach in precision measurement, offering unparalleled sensitivity beyond classical limits~\cite{RevModPhys.89.035002,PhysRevLett.123.213901,wang2021}. Recent advances exploit quantum coherence and entanglement to enable breakthroughs across diverse fields, neutrino mass measurements in particle physics~\cite{Bass2024}, nonlinear spectroscopic techniques via photon-pair correlations~\cite{Kutas2022}, optomechanical dark-matter detection in astrophysics, and superconducting circuit-based sensors~\cite{Danilin_2024}. Complementary developments in atomic systems, such as vapor-cell alkali atoms, diamond nitrogen-vacancy centers, and defects in silicon carbide, further expand the technological horizon~\cite{Cohen2020,ZhengzhiJiang2023,PhysRevLett.132.190001,PhysRevLett.127.186601}.

Enhancing sensor precision has emerged as a critical research focus in quantum sensing~\cite{PhysRevApplied.17.014034,PhysRevA.108.022215,Parto2025}. While Hermitian quantum systems can achieve sensitivity surpassing classical instruments at critical points, recent studies have demonstrated that non-Hermitian systems~\cite{Sarkar_2024,doi:10.1126/science.1258004,Wang023} offer a superior platform for quantum metrology~\cite{Ashida2020,PhysRevB.106.115107,WuYang2025,Ding2022}. By leveraging the critical dynamics of gain and loss in open quantum systems~\cite{McDonald2020,PhysRevLett.133.180801,PhysRevLett.133.180801}, these systems exhibit various precision scaling behaviors at critical or exceptional points~\cite{PhysRevA.108.053514,PhysRevA.102.033715,https://doi.org/10.1002/lpor.202100430}. In non-Hermitian sensors, researchers have conducted the self-consistent quantum analysis of the sensitivity and found that the nonreciprocity could effectively enhance sensing performance~\cite{Lau2018}. Meanwhile, the influence of noise in practical implementations warrants careful consideration~\cite{kcm2-2mz4,PhysRevLett.123.180501}. Despite the growing number of noise mitigation strategies~\cite{PhysRevX.8.021059,PhysRevLett.133.040801}, the inevitable presence of quantum noise and time constraints still poses significant challenges for quantum sensing. As a paradigmatic non-Hermitian system, parity-time (PT)-symmetric systems~\cite{PhysRevLett.113.053604,Peng2014,El-Ganainy2018,Feng2017,PhysRevApplied.19.034059,wang2025,RevModPhys.96.045002} have attracted significant attention due to their exceptional points (EPs)~\cite{doi:10.1126/science.aar7709,Chen2017,Wen_2018,GUO20243491} inducing eigenvalue degeneracy, which enables exponential sensitivity enhancement and facilitates the design of EP-based quantum sensors~\cite{PRXQuantum.6.020301,PhysRevA.111.052621,mondal2024,zhang2018,Wiersig2020}. Similarly, anti-PT symmetry~\cite{PhysRevResearch.4.033022} has recently emerged as a promising platform for quantum metrology~\cite{Peng2016,Wang24li}. Recent experimental demonstrations across diverse systems~\cite{Xu2023,Zhang2020,PhysRevLett.124.053901,10.1063/5.0020944} have confirmed the potential of anti-PT-symmetric criticality in achieving enhanced sensing performance~\cite{PhysRevLett.126.180401,PhysRevLett.129.273601,PhysRevApplied.13.014053,Wang2024,PhysRevB.105.064405}. The advent of dissipative coupling has significantly propelled advancements in quantum sensing~\cite{PhysRevLett.112.076402,s23218700}. This dissipative coupling mechanism, realized through buses or waveguides~\cite{10.1063/1.5144202,PhysRevB.99.134426}, enables long-range interactions among multiple modes. However, the impact of quantum noise and the development of dual-mode sensing architectures remain underexplored~\cite{HuLiang2025}.

In this work, we propose a system where two initially uncoupled modes interact through a dissipative coupling channel, demonstrating its potential for enhanced quantum sensing~\cite{PhysRevLett.126.180401,PhysRevLett.129.273601}. Unlike conventional PT-symmetric systems that rely on balanced gain and loss, we achieve the anti-PT symmetry by ensuring equal gain or loss in two independent modes through dissipative coupled channels. Using the quantum Langevin equation and the divergence properties of the transformation function, we derive the parameter relations and the probing frequency in the anti-PT symmetric broken/unbroken region. Furthermore, we analyze the quantum Cram\'er–Rao bound (QCRB) near the probing frequency by using the Laurent expansion approach. We find that when the frequency is close to the probing frequency, the behavior of the QCRB differs significantly across different phases. Such as, the QCRB may exhibit a deeper minimum near the EPs at the probing frequency in the broken phase. However, away from both the EPs and the probing frequency, the QCRB recovers a linear scaling behavior. In the unbroken phase, the QCRB consistently exhibits a linear relationship. In contrast to the above, the QCRB exhibits a quadratic dependence at the EPs due to its second-order nature, which indicates a pronounced enhancement in quantum sensing. This result is consistent with the conventional enhancement of sensing at EPs. Our scheme primarily relies on a shared environment to achieve dissipative coupling between two independent cavities, thereby constructing an anti-PT symmetric system. This proposal constructs the non-Hermitian Hamiltonian without relying on post-selection and is experimentally feasible, with potential implementations in platforms such as superconducting circuits and integrated photonic systems.

The rest of this paper is organized as follows: In Sec.~\ref{sec2}, we introduce a dissipatively coupled two-mode quantum system, and discuss its anti-PT symmetric properties through the Lindblad master equation framework. In Sec.~\ref{sec3}, the quantum Fisher information and QCRB are calculated near EPs. Then, Sec.~\ref{sec4} discusses the sensitivity across three distinct regions and includes a supplementary feasibility analysis. Finally, we provide a conclusion in Sec.~\ref{sec5}.

\section{Theoretical Model and Analysis}\label{sec2}
\begin{figure}[htpb]
\centering
\includegraphics[width=1\linewidth]{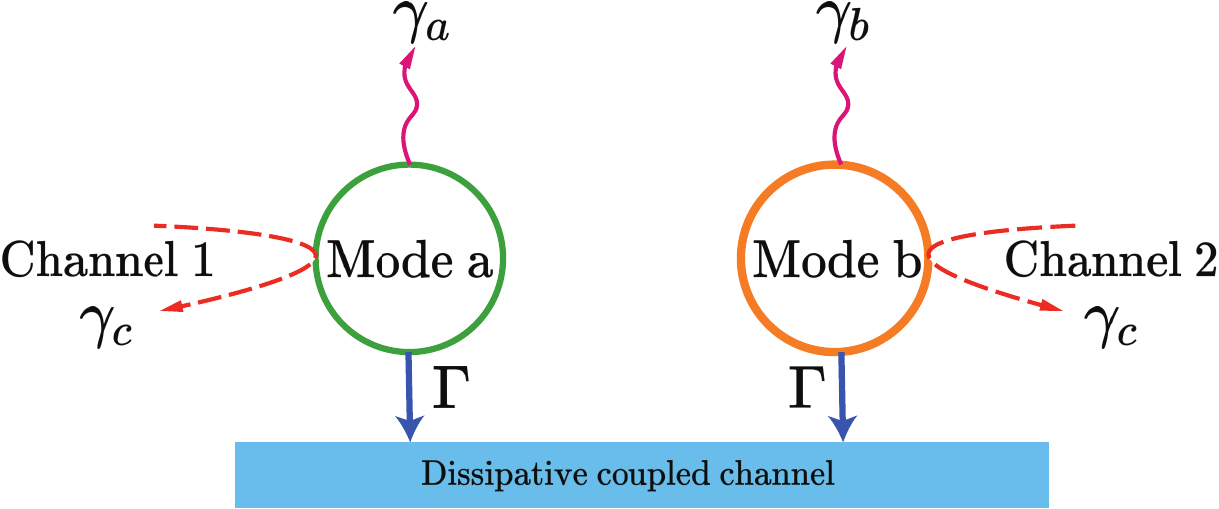}
\caption{Two separated modes ($a$ and $b$) interact with each other indirectly via the symmetric dissipation channel. Furthermore, each mode exhibits their intrinsic loss ($\gamma_{a}$ and $\gamma_{b}$) and the additive probe channel (Channels $1$ and $2$).}
\label{fig1}
\end{figure}

The studied model comprises two separated subsystems (mode $a$ and $b$), as illustrated in Fig.~\ref{fig1}, both coupled to a dissipative channel with a dissipation rate $\Gamma$~\cite{PhysRevLett.126.180401,PhysRevLett.129.273601}. This dissipation architecture induces a dissipative coupled channel alongside the common noise processes. In addition to the respective dissipative environment, each mode is also interacted with an independent probing channel with strength $\gamma_{c}$~\cite{Agarwal_2012}, enabling simultaneous readout and control of the system. The complete time evolution of the system's density matrix $\rho$ is governed by the following quantum master equation
\begin{eqnarray}
\frac{d\rho}{dt}=-\frac{i}{\hbar}[\hat{H},\rho]+\Gamma_a\mathcal{L}(\hat{a})\rho+\Gamma_b\mathcal{L}(\hat{b})\rho+2\Gamma\mathcal{L}(\hat{c})\rho,\label{eq1}
\end{eqnarray}
where all relevant interactions and environmental couplings have been incorporated. Here, $\hat{H}/\hbar=\omega_a\hat{a}^{\dagger}\hat{a}+\omega_b\hat{b}^{\dagger}\hat{b}$ is the system Hamiltonian of two spatially separated subsystems with resonance frequency $\omega_a$ and $\omega_b$, respectively. $\Gamma_a=\gamma_a+\gamma_c$ and $\Gamma_b=\gamma_b+\gamma_c$ represent the total dissipation rates of the modes $a$ and $b$, which include the intrinsic loss rate $\gamma_{a,b}$ and the coupling rate $\gamma_c$ with the probe channels. And we have assumed that the shared thermal bath is symmetrical to the two cavity modes and is written as $\hat{c}=(1/\sqrt{2})(\hat{a}+\hat{b})$. The Lindblad superoperator is defined by $\mathcal{L}(\hat{\sigma})\rho=2\hat{\sigma}\rho\hat{\sigma}^{\dagger}-\hat{\sigma}^{\dagger}\hat{\sigma}\rho-\rho\hat{\sigma}^{\dagger}\hat{\sigma}$ for any operator $\hat{\sigma}$.

\begin{figure}[htpb]
    \centering
    \includegraphics[width=1\linewidth]{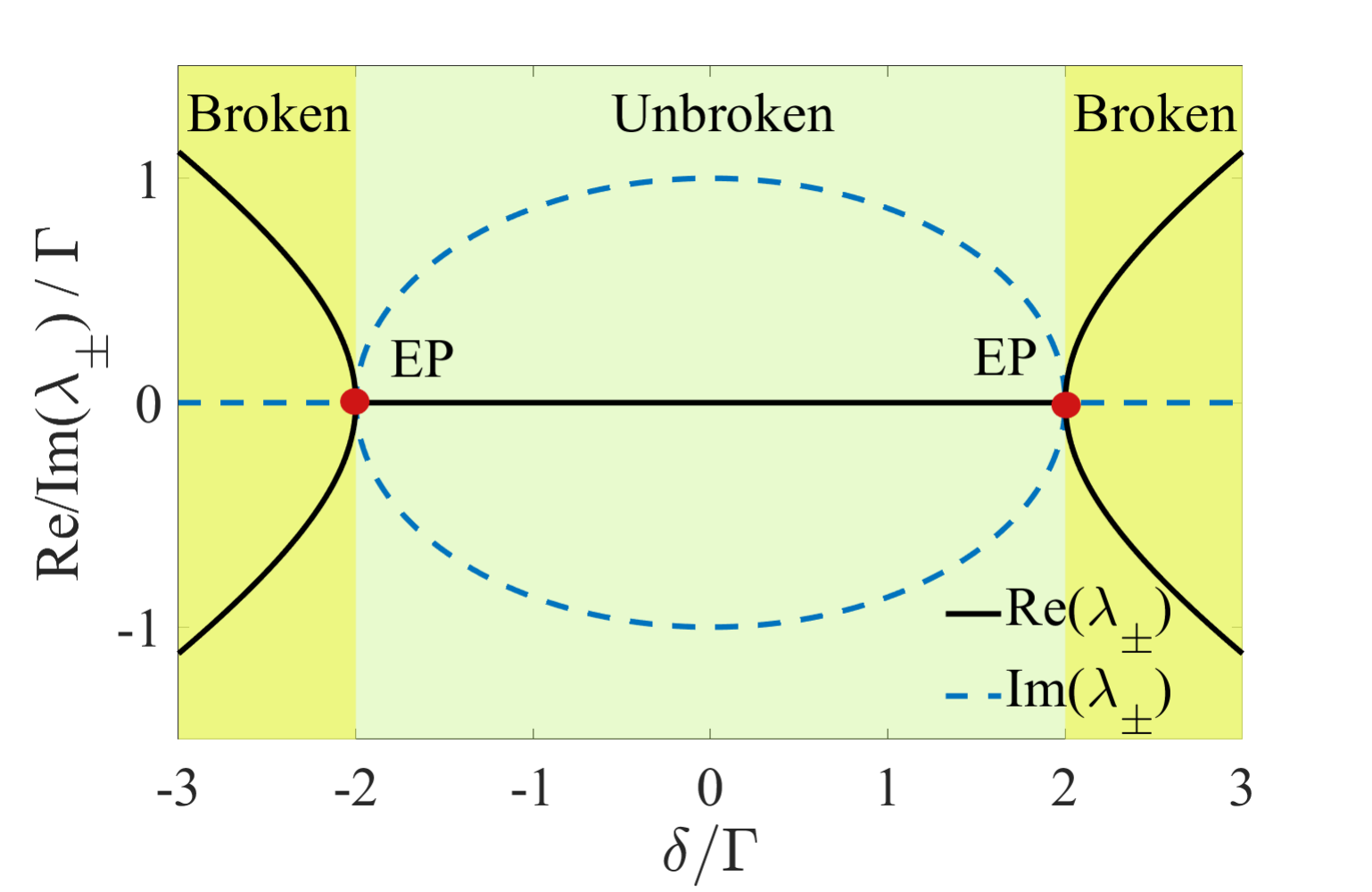}
    \caption{Eigenvalues $\lambda_{\pm}$. The black solid lines represent the real part of $\lambda_{\pm}$, and the blue dashed lines give the imaginary part of $\lambda_{\pm}$. The points at which the eigenvalues become degenerate are the EPs of the system, i.e., $\delta=\pm2\Gamma$. The region (green area) between two EPs corresponds to the unbroken anti-PT symmetric phase.}
    \label{fig2}
\end{figure}

The semiclassical dynamical equations for the cavity mode amplitudes $\langle \hat{a}\rangle$ and $\langle \hat{b}\rangle$ can be calculated by $\langle\dot{\hat{\sigma}}\rangle=\mathrm{Tr}[\hat{\sigma}\dot{\rho}]$ and expressed as
\begin{eqnarray}
\langle\dot{\hat{a}}\rangle&=&-\left(i\omega_{a}+\Gamma_{a}+\Gamma\right)\langle \hat{a}\rangle-\Gamma\langle \hat{b}\rangle,\cr\cr
\langle\dot{\hat{b}}\rangle&=&-\left(i\omega_{b}+\Gamma_{b}+\Gamma\right)\langle \hat{b}\rangle-\Gamma\langle \hat{a}\rangle.\label{eq2}
\end{eqnarray}
By analogy with the Schr\"odinger equation, the dynamical equations can be written in compact form $|\dot{\psi}\rangle=-i\mathcal{H}|\psi\rangle$ and $|\psi\rangle=(\langle \hat{a}\rangle,~\langle \hat{b}\rangle)^{\mathrm{T}}$. The effective non-Hermitian Hamiltonian is written as
\begin{eqnarray}
\mathcal{H}=
\begin{pmatrix}
\omega_a-i(\Gamma_a+\Gamma)&-i\Gamma\cr\cr
-i\Gamma&\omega_b-i(\Gamma_b+\Gamma)
\end{pmatrix}.\label{eq3}
\end{eqnarray}
In the rotating frame with frequency $\omega_0=(\omega_a+\omega_b)/2$, the Hamiltonian takes the simplified form
\begin{eqnarray}
\mathcal{H}_{\mathrm{eff}}=
\begin{pmatrix}
    \delta/2-i(\gamma_0+\Gamma)&-i\Gamma\cr\cr
    -i\Gamma&-\delta/2-i(\gamma_0+\Gamma)
\end{pmatrix},\label{eq4}
\end{eqnarray}
where $\delta=\omega_a-\omega_b$ is the detuning between the two cavities. And we have set $\gamma_0=\Gamma_a=\Gamma_b$ to balance the losses of the two cavities. It is easy to observe that the system satisfies the anti-PT symmetry $PT\mathcal{H}_{\mathrm{eff}}(PT)^{-1}=-\mathcal{H}_{\mathrm{eff}}$~\cite{PhysRevLett.126.180401,PhysRevApplied.13.014053,PhysRevResearch.7.023012}. In order to find the phase transition points under this condition, we calculate the characteristic equation $\mathrm{det}(\mathcal{H_{\mathrm{eff}}}-\lambda I)=0$. The eigenvalues and the eigenstates of the system are
\begin{eqnarray}\label{eq5}
    \lambda_{\pm}&=&-i(\gamma_0+\Gamma)\pm \sqrt{\frac{\delta^2}{4}-\Gamma^2},\cr\cr
    \psi_{\pm}&=&\left\{i\frac{\delta}{2\Gamma}\mp\sqrt{1-\frac{\delta^2}{4\Gamma^2}},1\right\}.
\end{eqnarray}
From Eq.~\eqref{eq5}, we can see that when condition $\delta=\pm2\Gamma$ is satisfied, the eigenvalues of the system become degenerate and eigenstates coalesce. This singular point is referred to as EPs in non-Hermitian systems, as shown in Fig.~\ref{fig2}. Under condition $|\delta|\leq2\Gamma$, the eigenvalues are purely imaginary and the system is in the unbroken anti-PT symmetric phase, whereas in other parameter regions, the anti-PT symmetry is spontaneously broken~\cite{PhysRevResearch.7.023012,Antinucci2024}. The presence of EP implies a sensing advantage in the system. So we will investigate the QCRB at this point and its vicinity in the following discussion.

\section{Quantum Cram\'er-Rao Bound and Sensitivity Enhancement Near Exceptional Points}\label{sec3}
In this dissipatively coupled two-cavity system, the quantum noise originates from multiple channels, such as the input noise through the probe port, the environmental coupling losses, and the noise induced by the dissipative coupling channel. Based on the effective non-Hermitian Hamiltonian $\mathcal{H_{\mathrm{eff}}}$ Eq.~\eqref{eq4}, we derive the quantum dynamical equations for the two cavity modes $a(b)$ via the quantum Langevin equation and the result is written as
\begin{eqnarray}
\dot{\hat{a}}&=&(-i\frac{\delta}{2}-\gamma_0-\Gamma)\hat{a}-\Gamma \hat{b}+\sqrt{\gamma_c}\hat{a}_{\mathrm{in}}+\sqrt{\gamma_{a}}\hat{N}_{a}+\sqrt{\gamma_{\Gamma}}\hat{N}^{\prime}_{a},\cr\cr
\dot{\hat{b}}&=&(i\frac{\delta}{2}-\gamma_0-\Gamma)\hat{b}-\Gamma \hat{a}+\sqrt{\gamma_c}\hat{b}_{\mathrm{in}}+\sqrt{\gamma_{b}}\hat{N}_{b}+\sqrt{\gamma_{\Gamma}}\hat{N}^{\prime}_{b},\label{eq6}
\end{eqnarray}
where $\hat{a}_{\mathrm{in}}(\hat{b}_{\mathrm{in}})$ denotes the input field in each probe channel. $\gamma_{i}=\gamma_{a/b}$ and $\gamma_{\Gamma}$ denote the intrinsic loss of each mode and the dissipative coupling rate, while $\hat{N}_{a/b}$ are the noise operators associated with intrinsic losses, the noise operators $\hat{N}^{\prime}_{a/b}$ arise from fluctuations in the dissipative coupling channel, modeled as vacuum noise.

To investigate the system’s response, the redefined quadrature components $\hat{x}_{a}=\hat{a}+\hat{a}^{\dagger}$, $\hat{y}_{a}=-i(\hat{a}-\hat{a}^{\dagger})$, $\hat{x}_{b}=\hat{b}+\hat{b}^{\dagger}$ and $\hat{y}_{b}=-i(\hat{b}-\hat{b}^{\dagger})$ are introduced. After performing the Fourier transformation defined by $\hat{a}(\omega)=\int\hat{a}(t)e^{-i\omega t}dt$, the Langevin equations are recast into a compact matrix form
\begin{eqnarray}
    &\begin{bmatrix}
        \hat{x}_a\\
        \hat{x}_b\\
        \hat{y}_a\\
        \hat{y}_b
    \end{bmatrix}&
    =\mathcal{G}_{APT}(\omega)\cr\cr
    &&\left(\sqrt{\gamma_c}\begin{bmatrix}
        \hat{x}_a^{\mathrm{in}}\\
        \hat{x}_b^{\mathrm{in}}\\
        \hat{y}_a^{\mathrm{in}}\\
        \hat{y}_b^{\mathrm{in}}
    \end{bmatrix}+\sqrt{\gamma_{i}}\begin{bmatrix}
    \hat{q}_{a}\\
    \hat{q}_{b}\\
    \hat{p}_{a}\\
    \hat{p}_{b}
    \end{bmatrix}+\sqrt{\gamma_{\Gamma}}\begin{bmatrix}
    \hat{q}^{\prime}_{a}\\
    \hat{q}^{\prime}_{b}\\
    \hat{p}^{\prime}_{a}\\
    \hat{p}^{\prime}_{b}
    \end{bmatrix}\right),\label{eq7}
\end{eqnarray}
where the transformation function $\mathcal{G}_{APT}(\omega)$ in Eq.~\eqref{eq7} is given by
\begin{eqnarray}
\mathcal{G}_{APT}(\omega)=
\begin{bmatrix}
    \gamma_0+\Gamma&\Gamma&\omega-\frac{\delta}{2}&0\\
    \Gamma&\gamma_0+\Gamma&0&\omega+\frac{\delta}{2}\\
    -\omega+\frac{\delta}{2}&0&\gamma_0+\Gamma&\Gamma\\
    0&-\omega-\frac{\delta}{2}&\Gamma&\gamma_0+\Gamma
    \end{bmatrix}^{-1}.\label{eq8}
\end{eqnarray}
And the quadrature terms of the input signal and other noise terms in the above expression are expressed as
\begin{eqnarray}
&\hat{x}_{a}^{\mathrm{in}}=&\hat{a}_{\mathrm{in}}+{\hat{a}_{\mathrm{in}}}^{\dagger},~~~~~~~~~~\hat{y}_{a}^{\mathrm{in}}=-i(\hat{a}_{\mathrm{in}}-{\hat{a}_{\mathrm{in}}}^{\dagger}),\cr\cr
&\hat{x}_{b}^{\mathrm{in}}=&\hat{b}_{\mathrm{in}}+{\hat{b}_{\mathrm{in}}}^{\dagger},~~~~~~~~~~~\hat{y}_{b}^{\mathrm{in}}=-i(\hat{b}_{\mathrm{in}}-{\hat{b}_{\mathrm{in}}}^{\dagger}),\cr\cr
&\hat{q}_{a/b}=&\hat{N}_{a/b}+\hat{N}_{a/b}^{\dagger},~~~~\hat{p}_{a/b}=-i(\hat{N}_{a/b}-\hat{N}_{a/b}^{\dagger}),\cr\cr
&\hat{q}^{\prime}_{a/b}=&\hat{N}^{\prime}_{a/b}+\hat{N}^{\prime\dagger}_{a/b},~~~~\hat{p}^{\prime}_{a/b}=-i(\hat{N}^{\prime}_{a/b}-\hat{N}^{\prime\dagger}_{a/b}),
\label{eq9}
\end{eqnarray}
where $\hat{x}_{a/b}^{\mathrm{in}}$ and $\hat{y}_{a/b}^{\mathrm{in}}$ are the quadratures of the input fields, $\hat{q}_{a/b}$, $\hat{p}_{a/b}$, $\hat{q}^{\prime}_{a/b}$, and $\hat{p}^{\prime}_{a/b}$ represent the intrinsic losses of the two cavities and the orthogonal basis of noise induced by discrete coupling channels, respectively.

To calculate the quantum Fisher information of the output field of probe channel, we introduce the input-output relationship of the system $\mu_{\mathrm{out}}(\omega)=[\hat{x}_a^{\mathrm{out}},\hat{x}_b^{\mathrm{out}},\hat{y}_a^{\mathrm{out}},\hat{y}_b^{\mathrm{out}}]^T=[\hat{x}_a^{\mathrm{in}},\hat{x}_b^{\mathrm{in}},\hat{y}_a^{\mathrm{in}},\hat{y}_b^{\mathrm{in}}]^T-\sqrt{\gamma_{c}}[\hat{x}_a,\hat{x}_b,\hat{y}_a,\hat{y}_b]^T$. Meanwhile, the output signal of the system is given by
\begin{eqnarray}
\mu_{\mathrm{out}}(\omega)
&=&(I-\gamma_c\mathcal{G}_{APT})
\begin{bmatrix}
\hat{x}_a^{\mathrm{in}}\\
\hat{x}_b^{\mathrm{in}}\\
\hat{y}_a^{\mathrm{in}}\\
\hat{y}_b^{\mathrm{in}}
\end{bmatrix}
-\sqrt{\gamma_c\gamma_{i}}\mathcal{G}_{APT}
\begin{bmatrix}
\hat{q}_{a}\\
\hat{q}_{b}\\
\hat{p}_{a}\\
\hat{p}_{b}
\end{bmatrix}\cr\cr
&&-\sqrt{\gamma_c\gamma_{\Gamma}}\mathcal{G}_{APT}
\begin{bmatrix}
\hat{q}^{\prime}_{a}\\
\hat{q}^{\prime}_{b}\\
\hat{p}^{\prime}_{a}\\
\hat{p}^{\prime}_{b}
\end{bmatrix},\label{eq10}
\end{eqnarray}
where $I$ is the $4\times4$ identity matrix. The covariance matrix $V_{\mathrm{out}}(\omega)$ for the output noise is derived as
\begin{eqnarray}
V_{\mathrm{out}}(\omega)&=&\mu_{\mathrm{out}}\cdot\mu_{\mathrm{out}}^{T}=(I-\gamma_c\mathcal{G}_{APT})V_{\mathrm{in}}(I-\gamma_c\mathcal{G}_{APT})^{T}\cr\cr
&&+\gamma_c\gamma_{i}\mathcal{G}_{APT}V_{\mathrm{i}}\mathcal{G}_{APT}^{T}+\gamma_c\gamma_\Gamma\mathcal{G}_{APT}V_\Gamma\mathcal{G}_{APT}^{T},\label{eq11}
\end{eqnarray}
where $V_{\mathrm{in}}$, $V_{\mathrm{i}}$, and $V_{\Gamma}$ denote the covariance matrices of the input channel, intrinsic loss, and dissipative coupling channel, respectively. Assuming that both the probe channel and the shared reservoir are in the vacuum noise~\cite{RevModPhys.82.1155}, these covariance matrices of input noises take the form as
\begin{eqnarray}
V_{\mathrm{in}}=V_{\mathrm{i}}=V_{\Gamma}=\begin{bmatrix}
    1&0&i&0\\
    0&1&0&i\\
    -i&0&1&0\\
    0&-i&0&1
\end{bmatrix}.\label{eq12}
\end{eqnarray}
According to the covariance matrix of the quantum noise, the average values of the quantum noise ($\hat{q}_{a}$, $\hat{q}_{b}$, $\hat{q}^{\prime}_{a}$, $\hat{q}^{\prime}_{b}$) and ($\hat{p}_{a}$, $\hat{p}_{b}$, $\hat{p}^{\prime}_{a}$, $\hat{p}^{\prime}_{b}$) are zero. The noise terms of the environment and the dissipative coupling channel in the output matrix can be reduced to zero, and the output signal in Eq.~\eqref{eq11} can be written as
\begin{eqnarray}
\mu_{\mathrm{out}}(\omega)=(I-\gamma_c\mathcal{G}_{APT}(\omega))
\begin{bmatrix}
\hat{x}_a^{\mathrm{in}}\\
\hat{x}_b^{\mathrm{in}}\\
\hat{y}_a^{\mathrm{in}}\\
\hat{y}_b^{\mathrm{in}}
\end{bmatrix}.\label{eq13}
\end{eqnarray}
As the photonic process is governed by Gaussian statistics, the QFI can be directly determined from the average signal output and covariance matrix of the output signal~\cite{Safranek_2019}
\begin{eqnarray}
I(\omega)=(\frac{\mathrm{d}\mu_{\mathrm{out}}(\omega)}{\mathrm{d}\omega})^{T}V_{\mathrm{out}}^{-1}(\omega)(\frac{\mathrm{d}\mu_{\mathrm{out}}(\omega)}{\mathrm{d}\omega}).\label{eq14}
\end{eqnarray}

From Eqs.~(\ref{eq11},\ref{eq14}), we can know that QFI $I(\omega)$ is only related to the detection input channel $\mu_{\mathrm{in}}(\omega)$ and the transformation function $\mathcal{G}_{APT}(\omega)$.
To enforce this divergence, the inverse transformation function $\mathcal{G}^{-1}_{APT}(\omega)$ must become singular at the detection frequency $\omega$. This condition is mathematically expressed through the determinant equation
\begin{eqnarray}
\mathrm{det}[\mathcal{G}^{-1}_{APT}(\omega)]=\left|\begin{array}{cccc}
   \gamma_0+\Gamma&\Gamma&\omega-\frac{\delta}{2}&0\\
    \Gamma&\gamma_0+\Gamma&0&\omega+\frac{\delta}{2}\\
    -\omega+\frac{\delta}{2}&0&\gamma_0+\Gamma&\Gamma\\
    0&-\omega-\frac{\delta}{2}&\Gamma&\gamma_0+\Gamma
\end{array}
\right|=0.\label{eq15}
\cr\cr
\end{eqnarray}
By solving Eq.~\eqref{eq15}, we can give the critical frequency $\omega$ condition
\begin{eqnarray}
\omega_{\pm}^2&=&-(\gamma_0^2+2\gamma_0\Gamma+2\Gamma^2)+\frac{\delta^2}{4}\cr\cr
&&\pm\sqrt{(\gamma_0+\Gamma)^2(4\Gamma^2-\delta^2)}.\label{eq16}
\end{eqnarray}
To ensure the detection frequency $\omega$ remains purely real, we must constrain parameter $\omega_{\pm}^{2}\geq0$. 

\textit{Case I:} The system is in the anti-PT symmetric broken region $|\delta/\Gamma|>2$. When the condition $\gamma_0+\Gamma=0$ is satisfied (balancing intrinsic, detection, and dissipative loss), the admissible detection frequency is obtained
\begin{eqnarray}
\omega_{\pm}^{I}=\pm\sqrt{\frac{\delta^2}{4}-\Gamma^2}.\label{eq17}
\end{eqnarray}
It is shown that there exists a frequency which satisfies the divergence condition and corresponds to the lasing threshold of the system when the anti-PT phase is broken.

\begin{figure}[htpb]
\centering
\includegraphics[width=1\linewidth]{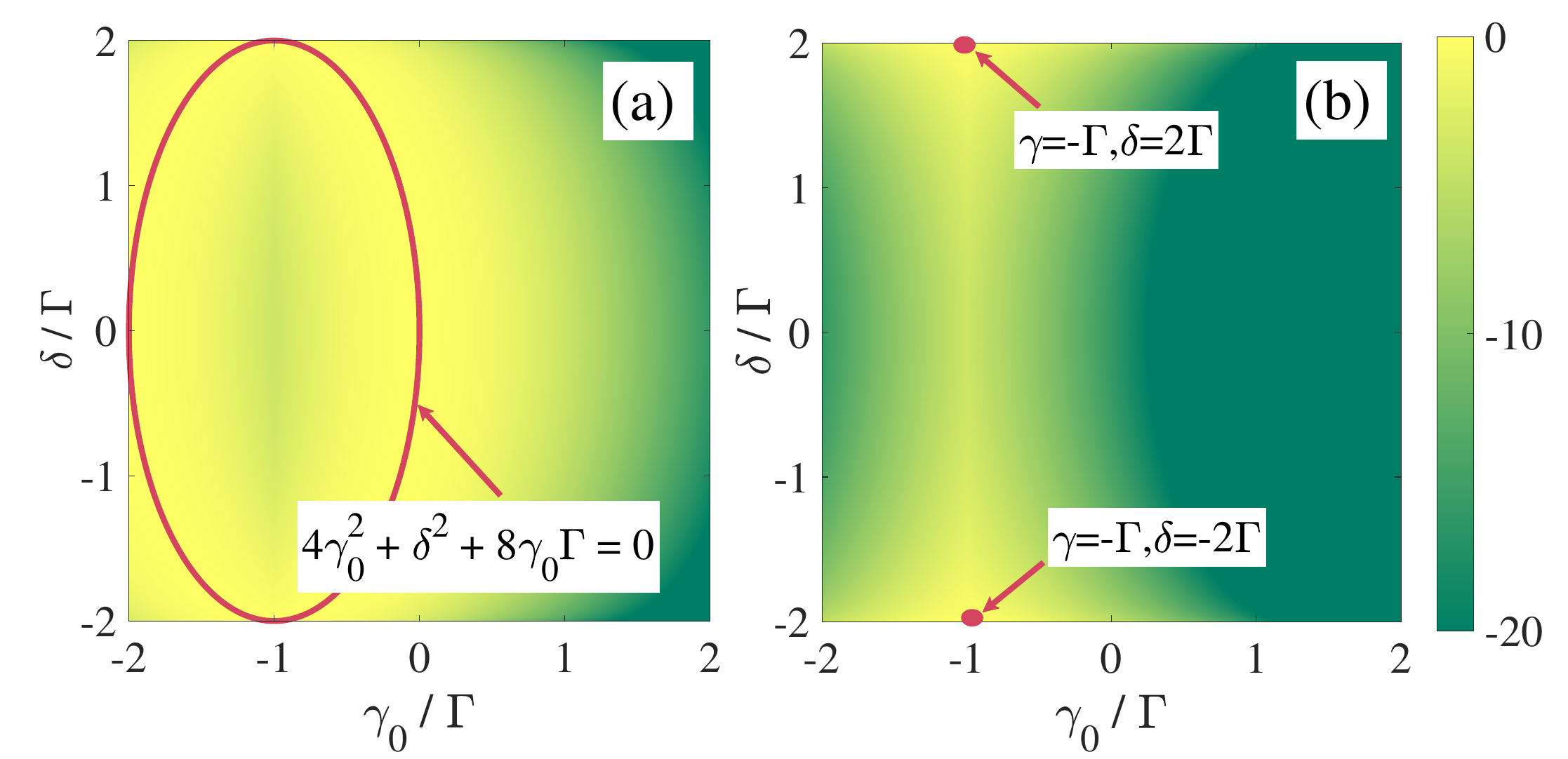}
\caption{The magnitude of $\omega_{\pm}^2$ in anti-PT symmetric unbroken region. (a) The positive branch of the expression for $\omega_{+}^2$, and the red elliptical line denotes the relationship in Eq.~\eqref{eq18}. (b) The negative branch of $\omega_{-}^2$, and the two red points are $\gamma_0/\Gamma=-1$ and $\delta/\Gamma=\pm 2$.}
\label{fig3}
\end{figure}

\textit{Case II:} When the system is in the anti-PT symmetric unbroken region $|\delta/\Gamma|\leq2$, the critical frequency must satisfies $\omega_{\pm}^{II}=0$. When $\omega_{+}^{II}=0$, we can get the follow relation
\begin{eqnarray}
4\gamma_0^2+\delta^2+8\gamma_0\Gamma=0.\label{eq18}
\end{eqnarray}
On the other hand, the balance condition need be satisfied $\gamma_0/\Gamma=-1$ and $\delta/\Gamma=\pm2$ for the $\omega_{-}^{II}=0$. The above results are shown in Fig.~\ref{fig3}. We can see that the maximum of the square of the critical frequency is 0 when the system is in the anti-PT symmetric unbroken phase.

Under this condition, in order to study the sensitivity at the detuned frequency, we perform the Laurent expansion~\cite{PhysRevLett.72.3439} on the transformation function at the frequency around $\omega_{\pm}^{I(II)}=\omega_0$
\begin{eqnarray}
\mathcal{G}_{APT}(\omega)&=&\sum_{k=0}^{m}C_k\times\frac{1}{(\omega-\omega_0)^k}\cr\cr
&\sim& C_m\times\frac{1}{(\omega-\omega_0)^m},\label{eq19}
\end{eqnarray}
where $C_{k/m}$ is the expanded coefficient matrix with the pole order $k(m)$. If the transformation function $\mathcal{G}_{APT}(\omega)$ can be expanded to Eq.~\eqref{eq19} around the defective point $\omega_0$, we can determine that the two parameters in QFI satisfy $V_{\mathrm{out}}^{-1}\sim (\omega-\omega_0)^{2m}$ and $d\mathcal{G}_{APT}(\omega)/d\omega\sim\mathcal{G}_{APT}^{2}\sim(\omega-\omega_0)^{-2m}$ based on Eq.~\eqref{eq11}. Then we know that the QFI is only related to the order of the transformation function $m$ and exhibits a relationship with $I(\omega)\sim(\omega-\omega_0)^{-2m}$.
We can determine the lower sensitivity limit $\delta\omega$ through the relationship between the QCRB~\cite{Liu2019,PhysRevA.97.042322,Rao:2008} and QFI $\delta\omega=1/\sqrt{I(\omega)}$. Similarly, we can also obtain QCRB that should meet the requirements as
\begin{eqnarray}
\delta\omega\sim(\omega-\omega_0)^m.\label{eq20}
\end{eqnarray}
We can characterize the sensitivity of the transformation function by expanding its series~\cite{PhysRevA.89.032128}. This scaling law highlights that the sensitivity near the EPs is critically enhanced by the pole order $m$ of the transformation function.

\section{Analysis and Discussion}\label{sec4}

To ensure that the frequency $\omega$ remains real-valued, we analyze both the anti-PT symmetric unbroken and broken regions. (i) In the unbroken region, satisfying condition $|\delta/\Gamma|\leq2$, there exists a continuous set of points satisfying the condition for Eq.~\eqref{eq18}, see the red line in Fig.~\ref{fig3}(a). This case corresponds to the undetuned lasing. (ii) Conversely, when the system is in $|\gamma_0/\Gamma|>2$ (anti-PT symmetric broken region) and $\gamma_0=-\Gamma$, the frequency condition is expressed in Eq.~\eqref{eq17}. (iii) The EPs $|\delta/\Gamma|=2$ are at the critical points. Next, we will discuss QCRB for each of the three cases.

\subsection{Symmetric unbroken region} 
For the anti-PT symmetric unbroken phase, we find that the purely real solutions for $\omega_{\pm}$ exist only when the condition in Eq.~\eqref{eq18} is satisfied. By further simplifying Eq.~\eqref{eq18}, we obtain the following condition for $\delta$ as
\begin{eqnarray}
\delta=\sqrt{-4\gamma_0(\gamma_0+2\Gamma)}.\label{eq21}
\end{eqnarray}
Combining Eq.~\eqref{eq8} and Eq.~\eqref{eq21}, we derive the inverse transformation function $\mathcal{G}^{-1}_{APT}(\omega)$ given by 
\begin{widetext}
\begin{eqnarray}
\mathcal{G}^{-1}_{APT}(\omega)=
\begin{bmatrix}
    \gamma_0+\Gamma&\Gamma&-\sqrt{-\gamma_0(\gamma_0+2\Gamma)}+\omega&0\\
    \Gamma&\gamma_0+\Gamma&0&\sqrt{-\gamma_0(\gamma_0+2\Gamma)}+\omega\\
    \sqrt{-\gamma_0(\gamma_0+2\Gamma)}-\omega&0&\gamma_0+\Gamma&\Gamma\\
    0&-\sqrt{-\gamma_0(\gamma_0+2\Gamma)}+\omega&\Gamma&\gamma_0+\Gamma
\end{bmatrix}.\label{eq22}
\end{eqnarray}
\end{widetext}
We perform a Laurent expansion of Eq.~\eqref{eq21} near $\omega_0=0$, expressing it in the form of Eq.~\eqref{eq19} as
\begin{eqnarray}
\mathcal{G}_{APT}(\omega)=
\begin{bmatrix}
h&0&-1&f\\
0&-h&f&-1\\
1&-f&h&0\\
-f&1&0&-h\\
\end{bmatrix}
\times\frac{1}{2\omega},\label{eq23}
\end{eqnarray}
where $h=\sqrt{-\gamma_0(\gamma_0+2\Gamma)}/(\gamma_0+\Gamma)$ and $f=\Gamma/(\gamma_0+\Gamma)$. The Laurent expansion in Eq.~\eqref{eq23} characterizes the QCRB as $\delta\omega\sim\omega$ at non-exceptional points with $\delta\neq\pm2\Gamma$ and $m=1$.

\subsection{Symmetric broken region} 
In the anti-PT symmetric broken phase, when the condition $\gamma_0=-\Gamma$ is satisfied (i.e., the gain-loss balance is achieved), a detunned lasing [see Eq.~\eqref{eq17}] emerges in the dissipatively coupled system. From Eq.~\eqref{eq8}, we derive the corresponding transformation function given by
\begin{eqnarray}
\mathcal{G}^{-1}_{APT}(\omega)=
\begin{bmatrix}
    0&\Gamma&\omega-\frac{\delta}{2}&0\\
    \Gamma&0&0&\omega+\frac{\delta}{2}\\
    -\omega+\frac{\delta}{2}&0&0&\Gamma\\
    0&-\omega-\frac{\delta}{2}&\Gamma&0
\end{bmatrix}.\label{eq24}
\end{eqnarray}
At the detuned frequency $\omega_{0}$, we perform the Laurent expansion of the transformation function according to Eq.~\eqref{eq24}, yielding as
\begin{eqnarray}
\mathcal{G}_{APT}=C_1\times \frac{1}{2\omega_0(\omega-\omega_0)}.\label{eq25}
\end{eqnarray}
For the case of $\omega_0=\sqrt{\delta^2/4-\Gamma^2}$ (valid for $\delta\neq\pm 2\Gamma$) and $C_1$ is the coefficient matrix encoding the system’s response to perturbations, which is defined as
\begin{eqnarray}
C_1=\begin{pmatrix}
    0&\Gamma&-\omega_0-\frac{\delta}{2}&0\\
    \Gamma&0&0&-\omega_0+\frac{\delta}{2}\\
    \omega_0+\frac{\delta}{2}&0&0&\Gamma\\
    0&\omega_0-\frac{\delta}{2}&\Gamma&0
\end{pmatrix}.\label{eq26}
\end{eqnarray}
The first-order pole ($m=1$) in the Laurent expansion Eq.~\eqref{eq25} directly dictates the scaling of the QCRB. Substituting Eq.~\eqref{eq25} into the QCRB expression Eq.~\eqref{eq20}, the frequency resolution scales as $\delta\omega \sim (\omega-\omega_0)$, indicating a linear divergence in sensitivity as $\omega\rightarrow\omega_0$.

\subsection{Exceptional Points}
The critical EPs between broken and unbroken phases satisfy the relation of $\delta=\pm2\Gamma$, enforcing the undetuned lasing frequency and the gain-loss balance. Under these dual conditions, the Laurent expansion of the transformation function at frequency $\omega_0=0$ reveals the dominant term of order as 
\begin{eqnarray}
\mathcal{G}_{APT}=C_2\times \frac{\Gamma}{\omega^2},\label{eq27}
\end{eqnarray}
where the symmetric coupling matrix $C_2$ is given by
\begin{eqnarray}
C_2=\begin{pmatrix}
    0&1&-1&0\\
    1&0&0&1\\
    1&0&0&1\\
    0&-1&1&0
\end{pmatrix}.\label{eq28}
\end{eqnarray}
We can see that the second-order pole in Eq.~\eqref{eq27} modifies the QCRB scaling to $\delta\omega\sim\omega^2$, indicating a quadratic divergence in sensitivity as $\omega\rightarrow0$. This enhancement comes from the defective Hamiltonian when the system is at the EPs, where the algebraic multiplicity of the Hamiltonian matrix is greater than its geometric multiplicity and it can amplify the system’s response to perturbations. It means that we can achieve the second-order enhancement of sensing sensitivity at the EPs, which is higher than the general points.

\begin{figure}[htpb]
    \centering
    \includegraphics[width=1\linewidth]{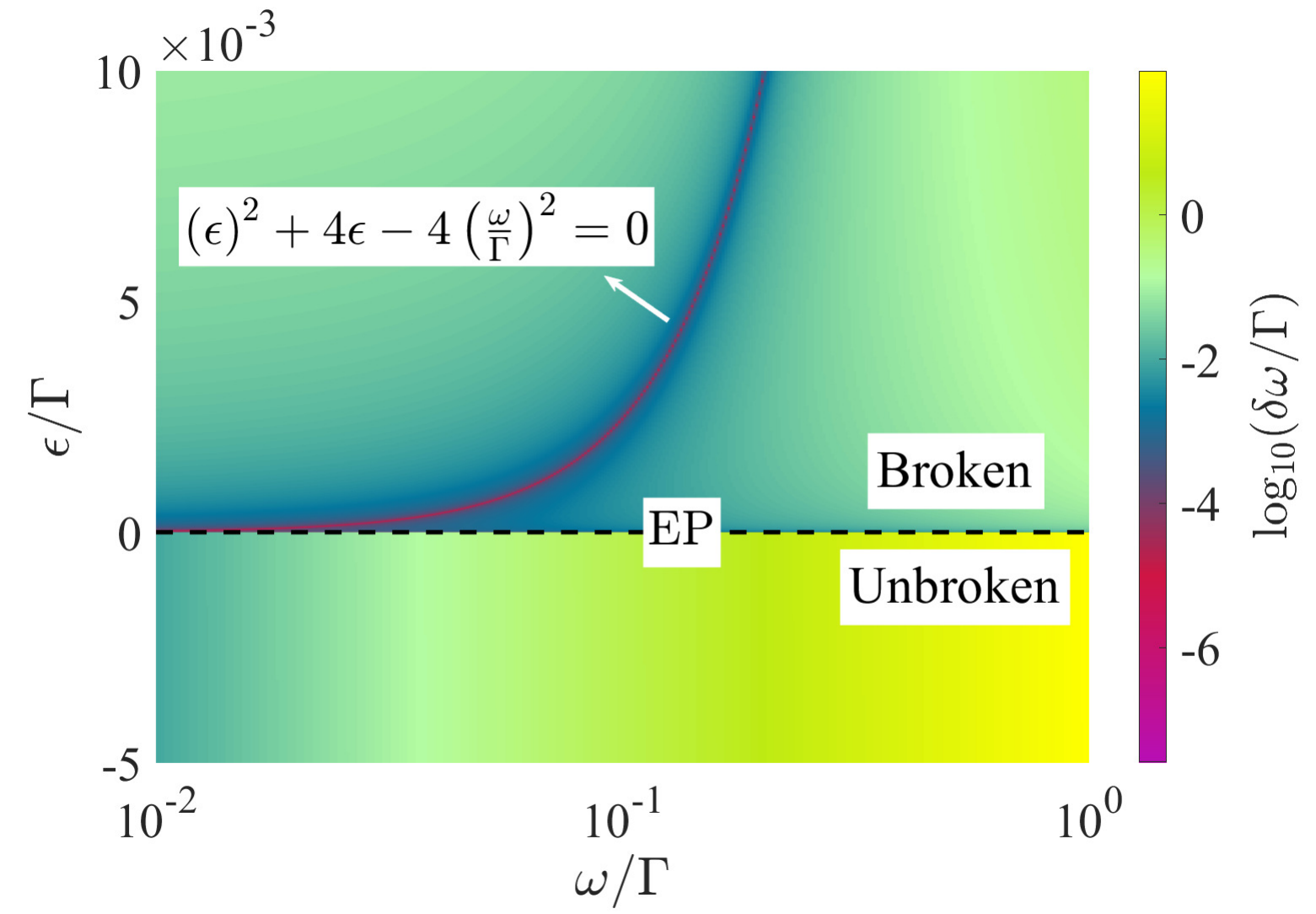}
    \caption{QCRB with a perturbation $\epsilon$. The black dashed line represents the system is at the EPs with $\epsilon=0$. The upper part with $\epsilon>0$ indicates the system is in the anti-PT symmetric broken phase. And the down part with $\epsilon<0$ indicates the system is in the anti-PT symmetric unbroken phase.}
    \label{fig4}
\end{figure}

To visualize the above analysis, we have shown the QCRB in Fig.~\ref{fig4} for different situations by adding a small perturbation on $\delta$, i.e., $\delta=(2+\epsilon)\Gamma$. Here, $\delta=2\Gamma$ represents the system is at the EPs, while the anti-PT symmetric broken and unbroken phases are discussed by setting the positive or negative perturbation symbol, respectively. We can clearly see that, if the perturbation is negative and the system is in the anti-PT symmetric unbroken region, the magnitude of QCRB is linearly related to $\omega$. For a small change of perturbation, the change of QCRB is not obvious, but the linear change relationship is always maintained. However, when the perturbation is positive (the system is in the anti-PT symmetric broken region), the change of QCRB with $\omega$ is no longer a linear relationship. There is a valley value when the detection frequency satisfies the relationship in Eq.~\eqref{eq17}, becoming $\epsilon^{2}+4\epsilon-4(\omega/\Gamma)^{2}=0$ when the perturbation is introduced. This indicates that the system exhibits a lower sensitivity limit at this frequency. It is due to the divergence of Eq.~\eqref{eq25} at the frequency, which decreases as the perturbation weakens. For the zero-perturbation case (EP state), we can know that there is a quadratic divergence in sensitivity as $\omega$ from Eq.~\eqref{eq27}. However, these features cannot be directly observed from the Fig.~\ref{fig4}. Moreover, the linear relationship in the broken phase is also not distinctly visible.

\begin{figure}[htpb]
\centering
\includegraphics[width=1\linewidth]{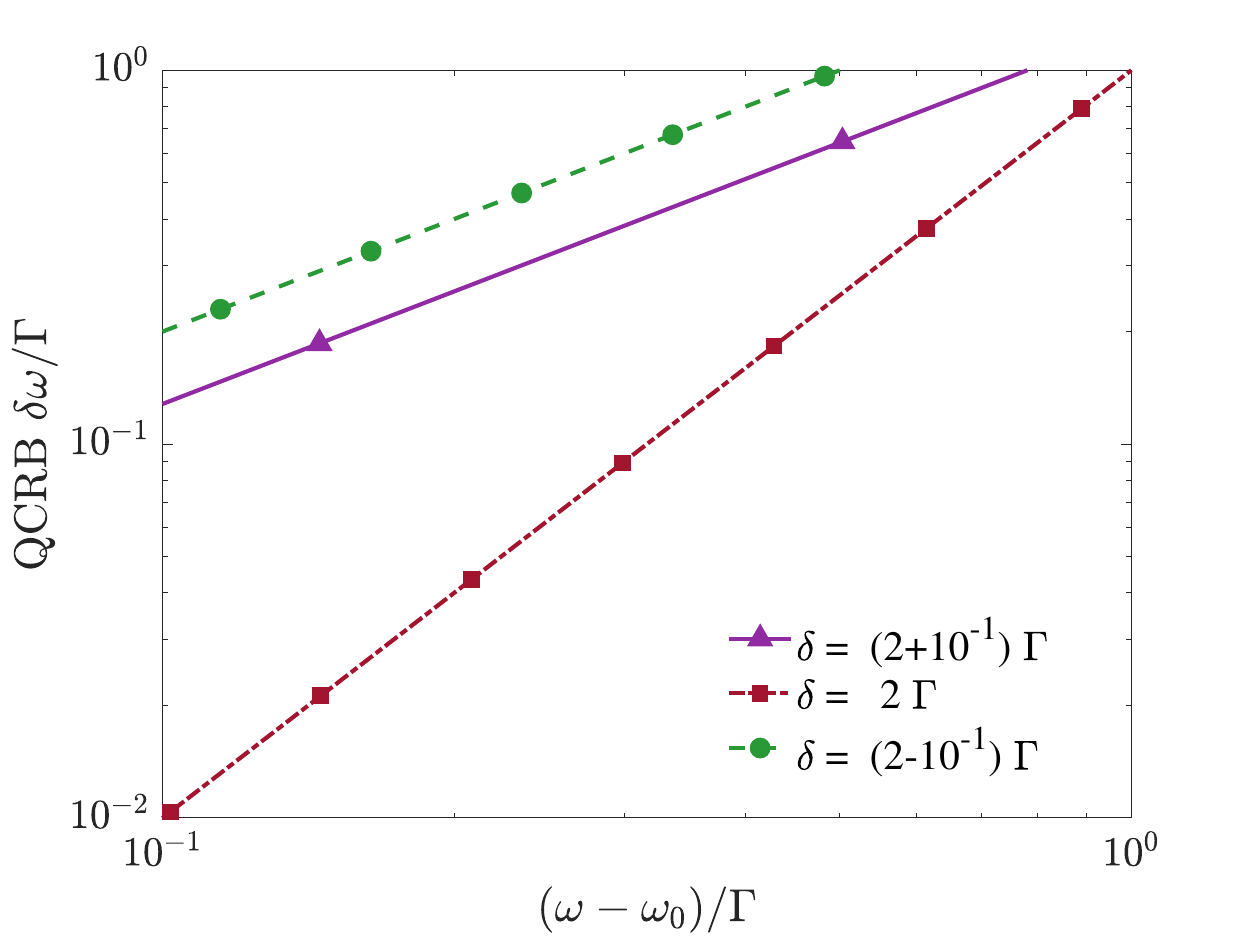}
\caption{QCRB in different situations. The solid line of purple triangle represents the case $\delta=(2+10^{-1})\Gamma$, the dotted line of red square corresponds to condition $\delta=2\Gamma$, and the green circle dotted line shows the result under condition $\delta=(2-10^{-1})\Gamma$.}
\label{fig5}
\end{figure}

To more intuitively illustrate the relationship between the QCRB and frequency, we plot the results with a larger perturbation ($5\%$) strength in Fig.~\ref{fig5} for the three different cases mentioned above. As can be observed, the detection sensitivity exhibits a linear dependence on the shifted probing frequency in logarithmic coordinates across all considered cases. However, the slope of this linear response varies across different cases. For instance, the slope at the EP is 2 (see the red dashed-dotted line in Fig.~\ref{fig5}), whereas it is 1 for the other two scenarios (see the green dashed line for the unbroken phase and the purple solid line for the broken phase). This implies that the QCRB displays quadratic scaling when the system is at the EPs, whereas in other regions it scales linearly. It is clearly demonstrated that the sensing at the EPs offers a superior advantage compared to operating in the regions on either side. The results are consistent with the conventional enhancement of sensing at the EPs of non-Hermitian system.

\section{Model Feasibility Analysis}
We construct an anti-PT symmetric system by introducing a dissipative coupling channel that interacts individually with two resonators, thereby establishing dissipative coupling between them. This channel, commonly referred to as a memory or bus in physical implementations~\cite{wang2025squeezingenhancedsensingexceptional}, has been widely adopted in various theoretical models and experimentally validated. Notably, such dissipative coupling is realizable not only in optical cavity systems~\cite{PhysRevA.110.052201,10.1063/5.0046202} but also in multimode platforms~\cite{PhysRevLett.124.053901} including optomechanical, superconducting qubit, and cavity magnonic systems.

\begin{figure}[htpb]
\centering
\includegraphics[width=1\linewidth]{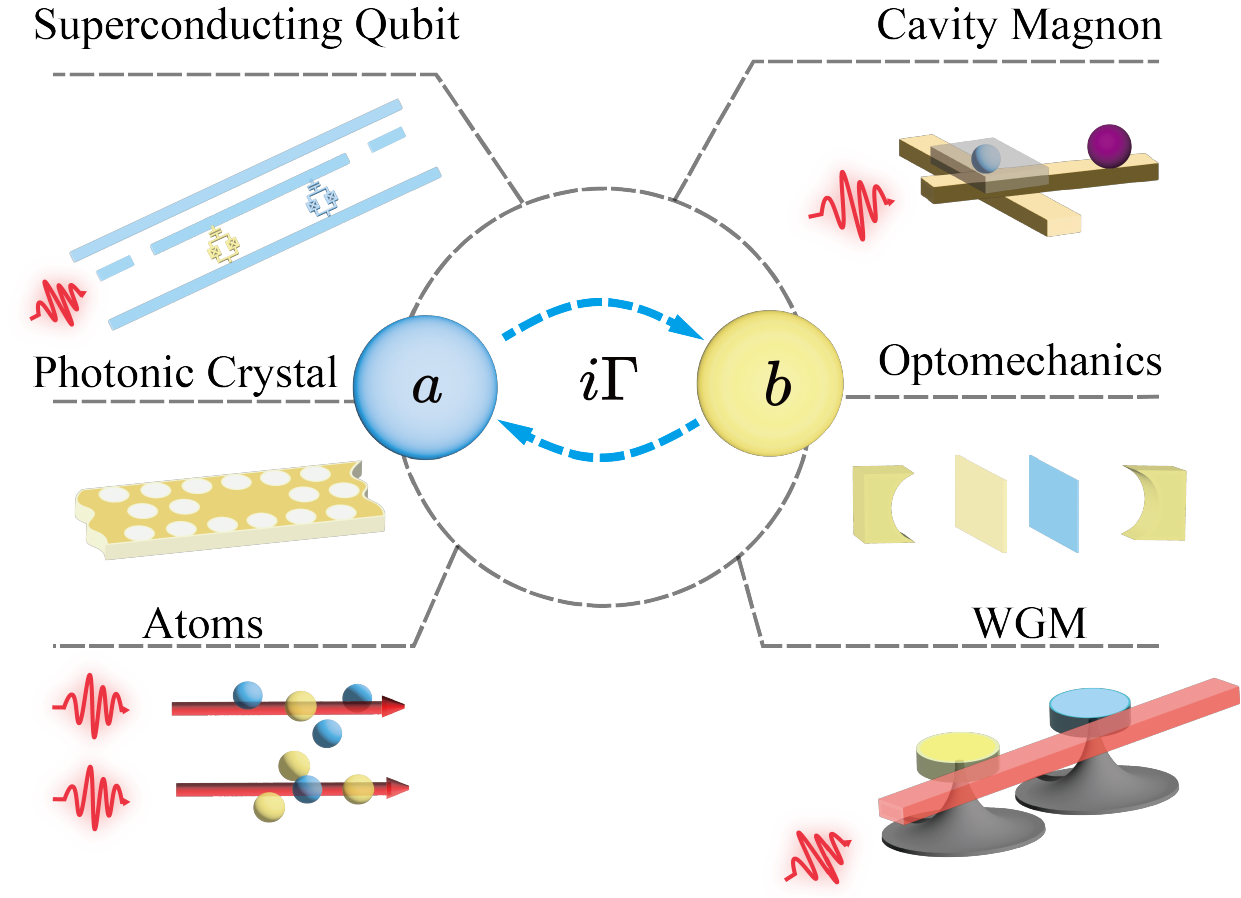}
\caption{Dissipative coupling realized in multiple systems. The system schematics: superconducting qubit~\cite{Majer2007}, cavity magnon~\cite{10.1063/1.5144202}, photonic crystal~\cite{Ellis2011}, optomechanics~\cite{PhysRevLett.124.053604}, atoms~\cite{Zhang2025}, whispering-gallery mode~\cite{PhysRevLett.132.193602}.}
\label{fig6}
\end{figure}

In studies of optomechanical systems~\cite{Xu2016,Gloppe2014}, a high-finesse optical cavity formed by a silicon nitride membrane serves as the reservoir. Experimentally, this square membrane ($1 ~\mathrm{mm} \times 1~\mathrm{mm} \times 50~\mathrm{nm}$) supports closed propagation paths. The reservoir effectively isolates coherent coupling between two optomechanical modes while providing resonant frequencies $\omega_{1}/(2\pi)=788.024~\mathrm{kHz}$ and $\omega_{2}/(2\pi)=788.487~\mathrm{kHz}$ for the two cavities. Without laser driving, the cavities remain nearly uncoupled and exhibit ultra-low decay rates $\gamma_{1/2}/(2\pi)\sim 1~\mathrm{Hz}$. Through precise control of detuning and driving power at $T=4.2~\mathrm{K}$, this configuration enables enhanced energy transfer at exceptional points.

For superconducting qubit systems, circuit quantum electrodynamics experiments have demonstrated quantum state transfer from qubits to photons~\cite{213q-sqxf,PhysRevB.106.L180406}. To achieve scalable architectures, early proposals introduce quantum busses for interqubit connectivity. In a two-qubit implementation, a microwave cavity operating at $\omega_{r}/(2\pi)=5.19~\mathrm{GHz}$ functions as a half-wave resonator ($L=\lambda/2=12.3~\mathrm{mm}$). When the transition frequency detuning between qubits significantly exceeds the coupling strength ($|\omega_{1}-\omega_{2}|\gg J$), their effective coupling is suppressed, enabling coherent state transfer. This long-range, low-loss interaction provides critical support for on-chip quantum information processing~\cite{Majer2007}. Multi-qubit extensions utilize transmission lines~\cite{PhysRevA.88.043806}, where individual qubits couple capacitively to the line, with excitation dynamics controlled through spatial arrangement. In electromechanical systems~\cite{Toth2017}, inductive coupling connects two circuits via a transmission line. The construction of bright and dark modes proves essential for satisfying the critical condition $\Gamma_{\mathrm{eff}}\gg k$.

Dissipative coupling through transmission lines has also been implemented in cavity magnonic systems~\cite{PhysRevLett.125.147202,10.1063/1.5144202,PhysRevLett.123.127202}. These configurations typically employ yttrium iron garnet spheres coupled to traveling waveguide modes under magnetic bias. Experimental realizations~\cite{PhysRevB.99.134426} operate at waveguide frequencies $\omega_{r}\sim~\mathrm{GHz}$. When the intrinsic losses of magnons and cavity photons satisfy $\kappa, \gamma \ll g$, the system's eigenvalues approach zero, ensuring information transfer predominantly through the waveguide. Beyond two-dimensional planar cavities, implementations using three-dimensional and quasi-one-dimensional cavities further confirm experimental feasibility.

The advantages of this model are also evident in experiments coupling optical and acoustic cavities~\cite{Zhu2024}. When a control laser and a signal laser are simultaneously injected into the optical cavity under phase-matching conditions, Brillouin scattering occurs, confining the interaction within a microsphere of approximately $90~\mathrm{\mu m}$. The Stimulated Brillouin Scattering coupling between the optical and acoustic cavities is activated by the detuning between the optical cavity's intrinsic frequency $\omega_{c0}$ and the control laser frequency $\omega_{c}$. The Brillouin scattering spectrum is obtained by scanning the signal laser frequency around $\omega_{s}=\omega_{c}-\Omega_{B}$, a condition corresponding to the parity-time-symmetric state, where $\Omega_{B}$ denotes the acoustic cavity frequency. This approach, benefiting from room-temperature operation and integrability, offers a promising platform for optical storage technologies.

In certain dual-cavity systems, the reservoir also plays a crucial role. When a common reservoir~\cite{PhysRevX.5.021025} is introduced alongside coherent cavity coupling, directional driving between cavities can be achieved provided the coherent and dissipative coupling strengths satisfy $J = i\Gamma/2$. Conversely, if a strongly dissipative auxiliary mode ($\kappa' \gg \kappa$), the resulting dissipation becomes non-directional and reciprocal. Such dissipative coupling can also be realized when both optical cavities simultaneously couple to a waveguide. In the taper-assisted coupling system~\cite{QIN2022128076}, a dual-bottle resonator and a microsphere resonator are separated by a distance of several micrometers ($\sim 3~\mathrm{mm}$). Dissipative coupling also exists between the two control beams of the same chirality, and its strength depends on the atomic diffusion rate~\cite{Zhang2025}. Dissipative coupling can also be realized in whispering-gallery mode resonators~\cite{PhysRevLett.132.193602,Kim2015}, either by having two cavities share a common transmission line, or by utilizing the interaction between counter-propagating optical modes in a single cavity. The spatial mode profile within a photonic crystal serves as a key mechanism that induces correlations between subsystems~\cite{Lodahl2017}.

\section{CONCLUSION} \label{sec5}
In conclusion, we analyze the sensing sensitivity in a dissipatively coupled general two-mode system, which exhibits the anti-PT symmetry under specific conditions. Our investigation covers the anti-PT symmetric unbroken region, the broken region, as well as the EPs and their vicinity. We find that in the unbroken region, only the undetuned critical frequency is present, and the QCRB varies linearly with the probing frequency. In the broken region, the detuned frequency emerges. When the probing frequency is near the detuned frequency, the QCRB decreases nonlinearly and has a valley value, enabling extremely high sensitivity within a narrow range. However, if the probing frequency is not sufficiently close to the detuned frequency, the behavior resembles that of the broken region. In contrast, when the system parameters meet the condition of EPs, the QCRB exhibits a quadratic dependence on the deviation of probing frequency, leading to superior and more robust sensing performance. Our work provides a theoretical foundation for noise-resilient sensing based on anti-PT symmetric systems and avoids the use of post-selection, opening avenues for the design of next-generation quantum sensors.

\begin{acknowledgments} 
This work was supported by the National Natural Science Foundation of China (Grants No. 12204424, No. 12147149, No. 12565001, No. 12205054, No. 12575032, No. 12274376), the China Postdoctoral Science Foundation (Grant No. 2022M722889), the Natural Science Foundation of Henan Province (Grant No. 232300421075), and the Natural Science Foundation of Jiangxi Province (No. 20252BAC200163).
\end{acknowledgments}

\section*{ DATA AVAILABILITY}
The data that support the findings of this article are openly
available~\cite{data}.

\bibliography{cite}
\end{document}